# Sector and Sphere: The Design and Implementation of a High Performance Data Cloud

Yunhong Gu
University of Illinois at Chicago

Robert L Grossman
University of Illinois at Chicago and Open Data Group

**ABSTRACT**

Cloud computing has demonstrated that processing very large datasets over commodity clusters can be done simply given the right programming model and infrastructure. In this paper, we describe the design and implementation of the Sector storage cloud and the Sphere compute cloud. In contrast to existing storage and compute clouds, Sector can manage data not only within a data center, but also across geographically distributed data centers. Similarly, the Sphere compute cloud supports User Defined Functions (UDF) over data both within a data center and across data centers. As a special case, MapReduce style programming can be implemented in Sphere by using a Map UDF followed by a Reduce UDF. We describe some experimental studies comparing Sector/Sphere and Hadoop using the Terasort Benchmark. In these studies, Sector is about twice as fast as Hadoop. Sector/Sphere is open source.

## 1. INTRODUCTION

By a cloud, we mean an infrastructure that provides on-demand resources or services over the Internet, usually at the scale and reliability of a data center. A storage cloud provides storage services (block or file-based services); a data cloud provides data management services (record-based, column-based or object-based services); and a compute cloud provides computational services. Often these are stacked together to serve as a computing platform for developing cloud-based applications.

Examples include Google's Google File System (GFS), BigTable and MapReduce infrastructure [8, 2, 4]; Amazon's S3 storage cloud, SimpleDB data cloud, and EC2 compute cloud [18]; and the open source Hadoop system [19], consisting of the Hadoop Distributed File System (HDFS), Hadoop's implementation of MapReduce, and HBase, an implementation of BigTable.

The implicit assumption with most high performance computing systems is that the processors are the scarce resource, and hence shared. When processors become available, the data is moved to the processors. To simplify, this is the supercomputing model. An alternative approach is to store the data and to co-locate the computation with the data when possible. To simplify, this is the data center model.

Cloud computing platforms (GFS/MapReduce/BigTable and Hadoop) that have been developed thus far have been designed with two important restrictions. First, clouds have assumed that all the nodes in the cloud are co-located, i.e., within one data center, or that there is relatively small bandwidth available between the geographically distributed clusters containing the data.

Second, these clouds have assumed that individual inputs and outputs to the cloud are relatively small, although the aggregate data managed and processed is very large. This makes sense since most clouds to date have targeted web applications in which large numbers of relatively small web pages are collected and processed as inputs, and outputs consist of search queries that return relatively small lists of relevant pages. Although some e-science applications have these characteristics, others must ingest relatively large datasets and process them. In addition, queries for certain e-science applications also result in relatively large datasets being returned.

In contrast, our assumption is that there are high speed networks (10 Gbps or higher) connecting various geographically distributed clusters and that the cloud must support both the ingestion and the return of relatively large data sets.

In this paper, we describe a storage cloud that we have developed called Sector and a compute cloud that we have developed called Sphere. Both of them are available as open source software from http://sector.sourceforge.net.

Sector is a distributed storage system that can be deployed over a wide area and allows users to ingest and download large datasets from any location with a high-speed network connection to the system. In addition, Sector automatically replicates files for better reliability, availability, and access throughout. Sector has been used to support distributing Sloan Digital Sky Survey data releases to astronomers around the world [12].

Sphere is a compute service built on top of Sector and provides a set of simple programming interfaces for users to write distributed data intensive applications. Sphere implements the stream processing paradigm, which is usually used in programming GPU [14] and multi-core processors. The stream processing paradigm can be used to implement any MapReduce supported applications.

The rest of this paper will describe the details of Sector and Sphere in Section 2 and Section 3, respectively. Section 4 describes some experimental studies. Section 5 describes related work and Section 6 is the summary and conclusion.

This is an expanded version of the conference paper [22]. This paper: a) describes a later version of Sector that includes security, b) includes additional information about how Sector works, including how security and scheduling are designed, and c) describes new experimental studies.

## 2. SECTOR

### 2.1 Overview

Sector is a storage cloud as defined above. Specifically, Sector provides storage services over the Internet with the scalability and reliability of a data center. Sector makes three assumptions:

1) First, Sector assumes that it has access to a large number of commodity computers (which we sometimes call nodes). The nodes may be located either within a data center or across data centers.

2) Second, Sector assumes that high-speed networks connect the various nodes in the system. For example, in the experimental studies described below, the nodes within a rack are connected by 1 Gbps networks, two racks within a data center are connected by 10 Gbps networks, and two different data centers are connected by 10 Gbps networks.

3) Third, Sector assumes that datasets it stores are divided into 1 or more separate files, which are called *Sector Slices*. The different files comprising a dataset are replicated and distributed over the various nodes managed by Sector. For example, one of the datasets managed by Sector in the experimental studies described below is a 1.3TB dataset consisting of 64 files, each approximately 20.3 GB in size.

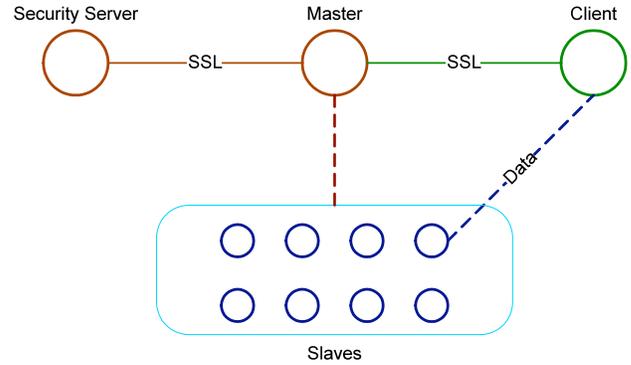

Figure 1. The Sector system architecture.

Figure 1 shows the overall architecture of the Sector system. The Security server maintains user accounts, user passwords, and file access information. It also maintains lists of IP addresses of the authorized slave nodes, so that illicit computers cannot join the system or send messages to interrupt the system.

The master server maintains the metadata of the files stored in the system, controls the running of all slave nodes, and responds to users' requests. The master server communicates with the security server to verify the slaves, the clients, and the users.

The slaves are the nodes that store the files managed by the system and process the data upon the request of a Sector client. The slaves are usually running on racks of computers that are located in one or more data centers.

### 2.2 File System Management

Sector is not a native file system; instead, it relies on the native file system on each slave node to store Sector Slices. A critical element in the design of Sector is that each Sector Slice is stored as one single file in the native file system. That is, Sector does not split Sector Slices into smaller chunks. This design decision greatly simplifies the Sector system and provides several advantages. First, with this approach, Sector can recover all the metadata it requires by simply scanning the data directories on each slave. Second, a Sector user can connect to a single slave node to upload or download a file. In contrast, if a storage cloud manages data at the block level, then a user will generally need to connect to many slaves to access all the blocks in a file. The Hadoop system is an example of a storage cloud that manages files at the block level [19]. Third, Sector can interoperate with native file systems if necessary.

A disadvantage of this approach is that it does require the user to break up large datasets into multiple files or to use a utility to accomplish this. Sector assumes that any user sophisticated enough to develop code for working with large datasets is sophisticated enough to split a large dataset into multiple files if required.

The master maintains the metadata index required by Sector and supports file system queries, such as file lookup and directory services.

The master also maintains the information about all slaves (e.g., available disk space) and the system topology, in order to choose slaves for better performance and resource usage.

The current implementation assumes that Sector will be installed on a hierarchical topology, e.g., computer nodes on racks within multiple data centers. The topology is manually specified by a configuration file on the master server.

The master checks the number of copies of each file periodically. If the number is below a threshold (the current default is 3), the master chooses a slave to make a new copy of the file. The new location of the file copy is based on the topology of the slaves' network. When a client requests a file, the master can choose a slave (that contains a copy of the file) that is close to the client and is not busy with other services.

The Sector client supports standard file access APIs, such as open(), read(), and write(), These APIs can be wrapped to support other standards, such as SAGA.

## 2.3 Security

Sector runs an independent security server. This design allows different security service providers to be deployed (for example LDAP and Kerberos). In addition, multiple Sector masters (for better reliability and availability) can use the same security service.

A client logs onto the master server via an SSL connection. The user name and the password are sent to the master. The master then sets up an SSL connection to the Security server and asks to verify the credibility of the client. The security server checks its user database and sends the result back to the master, along with a unique session ID and the client's file access privileges (for example its I/O permissions for different directories). In addition to the password, the client IP address is checked against an access control list defined for the user. Both SSL connections require the use of public certificates for verification.

If the client requests access to a file, the master will check whether the user has access privileges for that file. If granted, the master chooses a slave node to serve the client. The slave and the client then set up an exclusive data connection that is coordinated by the master. Currently, the data connection is not encrypted, but we expect to add encryption in a future release.

Sector slave nodes only accept commands from the Sector master. Neither Sector clients nor other slave nodes can send commands directly to a slave. All client-slave and slave-slave data transfer must be coordinated by the master node.

Finally, the security server controls whether a slave can be added to the system. The security server maintains an IP list and/or an IP range so that only computers on this list can join as slaves.

## 2.4 Message and Data Transfer

Sector uses UDP for message passing and UDT [11] for data transfer. UDP is faster than TCP for message passing because it does not require connection setup. We developed a reliable message passing library called GMP (Group Messaging Protocol) to use in Sector. For data transfer, a Sector slave will set up a UDT connection with the client directly. This UDT connection is set up using a rendezvous connection mode and is coordinated by the master. UDT is a high performance data transfer protocol and significantly outperforms TCP over long distance high bandwidth links [11].

A single UDP port is used for messaging and another single UDP port is used for all the data connections. A limited number of threads process the UDP packets, independent of the number of connections, which makes the communication mechanism scale nicely as the number of nodes in the system increases.

## 3. SPHERE

## 3.1 Overview

Recall that Sphere is a compute cloud that is layered over Sector. To introduce Sphere, consider the following example application. Assume we have 1 billion astronomical images of the universe from the Sloan Digital Sky Survey (SDSS) and the goal is to find brown dwarfs (a stellar object) in these images. Suppose the average size of an image is 1 MB so that the total data size is 1 TB. The SDSS dataset is stored in 64 files, named SDSS1.dat, …, SDSS64.dat, each containing one or more images.

In order to access an image randomly in the dataset (consisting of 64 files), we built an index file for each file. The index file indicates the start and end positions (i.e., offset and size) of each record (in this case, an image) in the data file. The index files are named by adding an ".idx" postfix to the data file name: SDSS1.dat.idx, …, SDSS60.dat.idx.

To use Sphere, the user writes a function "findBrownDwarf" to find brown dwarfs from each image. In this function, the input is an image, while the output indicates the brown dwarfs.

```
findBrownDwarf(input, output);
```

A standard serial program might look like this:

```
for each file F in (SDSS slices)
   for each image I in F
       findBrownDwarf(I, …);
```

Using the Sphere client API, the corresponding pseudo code looks like this:

```
SphereStream sdss;
sdss.init(/*list of SDSS slices*/);
SphereProcess myproc;
Myproc.run(sdss, "findBrownDwarf");
Myproc.read(result);
```

In the pseudo code fragment above, "sdss" is a Sector stream data structure that stores the metadata of the Sector slice files. The application can initialize the stream by giving it a list of file names. Sphere automatically retrieves the metadata from the Sector network. The last three lines will simply start the job and wait for the result using a small number of Sphere APIs. There is no need for users to explicitly locate and move data, nor do they need to take care of message passing, scheduling, and fault tolerance.

## 3.2 The Computing Paradigm

As illustrated in the example above, Sphere uses a stream processing computing paradigm. Stream processing is one of the most common ways that GPU and multi-core processors are programmed. In Sphere, each slave processor is regarded as an ALU in a GPU, or a processing core in a CPU. In the stream processing paradigm, each element in the input data array is processed independently by the same processing function using multiple computing units. This paradigm is also called SPMD (single program multiple data), a term derived from the Flynn's taxonomy of SIMD (single instruction multiple data) for CPU design.

We begin by explaining the key abstractions used in Sphere. Recall that a Sector dataset consists of one or more physical files. A *stream* is an abstraction in Sphere and it represents either a dataset or a part of a dataset. Sphere takes streams as inputs and produces streams as outputs. A Sphere stream consists of multiple data *segments* and the segments are processed by Sphere Processing Engines (SPEs) using slaves. An SPE can process a single data record from a segment, a group of data records, or the complete segment.

Figure 2 illustrates how Sphere processes the segments in a stream. Usually there are many more segments than SPEs, which provides a simple mechanism for load balancing, since a slow SPE simply processes fewer segments. Each SPE takes a segment from a stream as an input and produces a segment of a stream as output.

These output segments can in turn be the input segments to another Sphere process. For example, a sample function can be applied to the input stream and the resulting sample can be processed by another Sphere process.

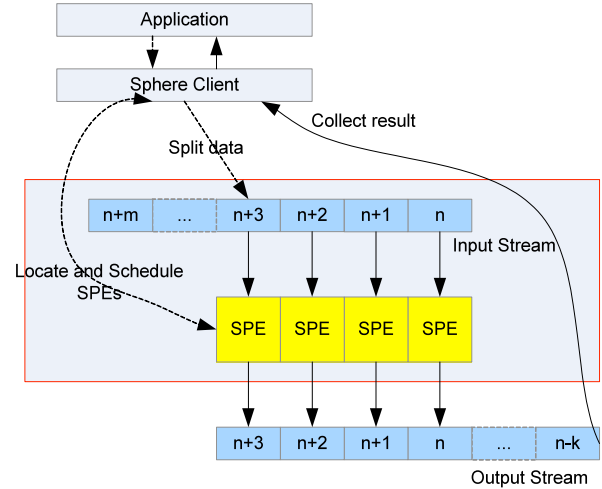

**Figure 2. The computing paradigm of Sphere.**

Figure 2 illustrates the basic model that Sphere supports. Sphere also supports some extensions of this model, which occur quite frequently.

**Processing multiple input streams.** First, multiple input streams can be processed at the same time (for example, the operation A[] + B[] is supported). Note that this is not a straightforward extension, because it can be complex to split input streams and to assign segments to SPEs.

**Shuffling input streams.** Second, the output can be sent to multiple locations, rather than just writing to local disk. Sometimes this is called shuffling. For example, a user-defined function can specify a bucket ID (that refers to a destination file on either a local or on a remote node) for each record in the output, and Sphere will send this record to the specified destination. At the destination, Sphere receives results from many SPEs and writes them into a file, in the same order that they arrive. It is in this way that Sphere supports MapReduce style computations [4].

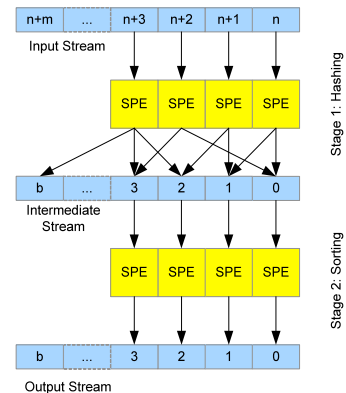

**Figure 3: Sorting large distributed datasets with Sphere**

Figure 3 shows an example that uses two Sphere processes (each process is called a stage) to implement distributed sorting. The first stage hashes the input data into multiple buckets. The hashing function scans the complete stream and places each element in a proper bucket. For example, if the data to be sorted is a collection of integers, the hashing function can place all data less than $T_0$ in bucket $B_0$, data between $T_0$ and $T_1$ in bucket $B_1$, and so on.

In stage 2, each bucket (which is a data segment) is sorted by an SPE. Note that after stage 2, the entire dataset (stream) is now sorted. This is because all elements in a bucket are smaller than all the elements in any buckets further along in the stream.

Note that in stage 2, the SPE sorts the *whole* data segment and does not just process each record individually.

**SPE can process records or collections of records.** This is the third expansion to the basic model. In Sphere, an SPE can process a single record, multiple records, the whole segment, or a complete file at one time.

### 3.3 Sphere Processing Engine

Once the master accepts the client's request for Sphere data processing, it sends a list of available slave nodes to the client. The client then chooses some or all the slaves and requests that an SPE start on these nodes. The client then sets up a UDT connection (for both control and data) with the SPE. The stream processing functions, in the form of dynamic libraries are sent to each SPE and stored locally on the slave node. The SPE then opens the dynamic libraries and obtains the various processing functions. Then it runs in a loop that consists of the following four steps:

First, the SPE accepts a new data segment from the client containing the file name, offset, number of rows to be processed, and various additional parameters.

Next, the SPE reads the data segment (and the corresponding portion of the idx index file if it is available) from either the local disk or from another slave node.

As required, the stream processing function processes either a single data record, a group of data records, or the entire segment, and writes the result to the proper destinations. In addition, the SPE periodically sends acknowledgments to the client about the progress of the current processing.

When the data segment is completely processed, the SPE sends an acknowledgment to the client to conclude the processing of the current data segment.

If there are no more data segments to be processed, the client closes the connection to the SPE, and the SPE is released. The SPE may also timeout if the client is interrupted.

### 3.4 Sphere Client

The Sphere client provides a set of APIs that developers can use to write distributed applications. Developers can use these APIs to initialize input streams, upload processing function libraries, start Sphere processes, and read the processing results.

The client splits the input stream into multiple data segments, so that each can be processed independently by an SPE. The SPE can either write the result to the local disk and return the status of the processing or it can return the result of the processing itself. The client tracks the status of each segment (for example whether the segment has been processed) and holds the results.

The client is responsible for orchestrating the complete running of each Sphere process. One of the design principles of the Sector/Sphere system is to leave most of the decision making to the client so that the Sector master can be quite simple. In Sphere, the client is responsible for the control and scheduling of the program execution.

### 3.5 Scheduler

#### 3.5.1 Data Segmentation and SPE Initialization

The client first locates the data files in the input stream from Sector. If the input stream is the output stream of a previous stage, then this information is already within the Sector stream structure and no further segmentation is needed.

Both the total data size and the total number of records are calculated in order to split the data into segments. This is based on the metadata of the data files retrieved from Sector.

The client tries to uniformly distribute the input stream to the available SPEs by calculating the average data size per SPE. However, in consideration of the physical memory available per SPE and the data communication overhead per transaction, Sphere limits the data segment size between size boundaries $S_{min}$ and $S_{max}$ (the default values are 8MB and 128MB, respectively, but user defined values are supported). In addition, the scheduler rounds the segment size to a whole number of records since a record cannot be split. The scheduler also requires that a data segment only contain records from a single data file.

As a special case, the application may request that each data file be processed as a single segment. This would be the case, for example, if an existing application were designed to only process files. This is also the way the scheduler works when there is no record index associated with the data files.

*3.5.2 SPE Scheduling*

Once the input stream is segmented, the client assigns each segment to an SPE. The following rules are applied:

1. Each data segment is assigned to an SPE on the same node if there is one available.
2. Segments from the same file are processed at the same time unless following this rule leaves SPEs idle.
3. If there are still idle SPEs available after rule 1 and rule 2 are applied, assign them parts of data segments to process in the same order as they occur in the input stream.

The first rule tries to run an SPE on the same node that the data resides (in other words, to exploit data locality). This reduces the network traffic and yields better throughput. The second rule improves data access concurrency because SPEs can read data from multiple files independently at the same time.

As mentioned in Section 3.3, SPEs periodically provide feedback about the progress of the processing. If an SPE does not provide any feedback about the progress of the processing before a timeout occurs, then the client discards the SPE. The segment being processed by the discarded SPE is assigned to another SPE, if one is available, or placed back into the pool of unassigned segments. This is the mechanism that Sphere uses to provide fault tolerance. Sphere does not use any check pointing in an SPE; when the processing of a data segment fails, it is completely re-processed by another SPE.

Fault tolerance is more complicated when SPEs write results to multiple destinations (as happens when using buckets for example). Each SPE dumps the result to local disk before attempting to send the results to buckets on other nodes. In this way, if one node is down, the result can be sent to the same buckets on other nodes. Each bucket handler also records the status of incoming results from each data segment; thus if one SPE is down, the bucket handler can continue to accept data in the correct order from another SPE that processes the same data segment again.

If errors occur during the processing of a data segment due to problems with the input data or bugs in user defined functions, the data segment will not be processed by any other SPE. Instead, an error report is sent back to the client, so that the application can take the appropriate action.

In most cases, the number of data segments is significantly greater than the number of SPEs. For example, hundreds of machines might be used to process terabytes of data. As a consequence, the system is naturally load balanced because all SPEs are kept busy during the majority of the runtime. Imbalances only occur towards the end of the computation when there are fewer and fewer data segments to process, causing some SPEs to be idle.

Different SPEs can require different times to process data segments. There are several reasons for this, including: the slave nodes may not be dedicated; the slaves may have different hardware configurations (Sector systems can be heterogeneous); and different data segments may require different processing times. Near the end of the computation, when there are idle SPEs but incomplete data segments, each idle SPE is assigned one of the incomplete segments. That is, the remaining segments are run on more than one SPE and the client collects results from whichever SPE finishes first. In this way, Sphere avoids waiting for the slow SPEs while the faster ones are idle. After processing is complete, the Sphere client can reorder the segments to correspond to the original order in the input stream.

## 3.6 Comparison with MapReduce

Both the stream processing framework used by Sphere and the MapReduce framework can be viewed as ways to simplify parallel programming. The approach of applying User Defined Functions (UDF) to segments managed by a storage cloud is more general than the MapReduce approach in the sense that with Sphere it is easy to specify a Map UDF and to follow it with a reduce UDF. We now describe how to do this in more detail.

A MapReduce map process can be expressed directly by a Sphere process that writes the output stream to local storage. A MapReduce reduce process can be simulated by the hashing/bucket process of Sphere. In MapReduce, there is no data exchange between slave nodes in the map phase, while each reducer in the reduce phase reads data from all the slaves. In Sphere's version, the first stage hashes (key, value) pairs to buckets on other slave nodes, while in the second stage all data is processed locally at the slave by the reduce function.

We illustrate this by showing how MapReduce and Sphere compute an inverted index for a collection of web pages. Recall that the input is a collection of web pages containing terms (words) and the output is a sorted list of pairs (w, <pages>), where w is a word that occurs in the collection and <pages> is a list of web pages that contain the word w. The list is sorted on the first component.

Computing an inverted index using Sphere requires two stages. In the first stage, each web page is read, the terms are extracted, and each term is hashed into a different bucket. Sphere automatically assigns each bucket to a separate slave for processing. Think of this as the hashing or shuffling stage. To be more concrete, all words starting with the letter "a" can be assigned to the bucket 0, those beginning with the letter "b" to the bucket 1, and so on.  A more advanced hashing technique that would distribute the words more evenly could also be used.  In the second stage, each bucket is processed independently by the slave node, which generates a portion of the inverted index.  The inverted index consists of multiple files managed by Sector.

For example, assume that there are two web pages (each is a separate file): *w1.html* and *w2.html*.  Assume that *w1* contains the words "*bee*" and "*cow*" and that *w2* contains the words "*bee*" and "*camel*". In the first stage of Sphere, bucket 1 will contain (*bee*, *w1*) and (*bee w2*), and bucket 2 will contain (*cow*, *w1*) and (*camel*, *w2*).  In the second stage, each bucket is processed separately. Bucket 1 becomes (*bee*, (*w1*, *w2*)) and bucket 2 remains unchanged. In this way the inverted index is computed and the result is stored in multiple files (bucket files).

In Hadoop's MapReduce [19], the Map phase would generate four intermediate files containing (*bee*, *w1*), (*cow*, *w1*), (*bee*, *w2*), and (*camel*, *w2*). In the Reduce phase, the reducer will merge the same keys and generate three items (*bee*, (*w1*, *w2*)), (*cow*, *w1*), and (*camel*, *w2*).

## 4. EXPERIMENTAL STUDIES

We have released Sector/Sphere as open source software and used it in a variety of applications.  We have also analyzed its performance using the Terasort benchmark [9, 23].  In this section we describe two Sector/Sphere applications and discuss their performance.

### 4.1 SDSS Data Distribution

Sector is currently used to distribute the data products from the Sloan Digital Sky Survey (SDSS) over the Teraflow Testbed [12, 21]. We set up multiple Sector servers on the Teraflow Testbed that we use to store the SDSS data.  We stored the 13 TB SDSS Data Release 5 (DR5), which contains 60 catalog files, 64 catalog files in EFG format, and 257 raw image data collection files.  We also stored the 14 TB SDSS Data Release 6 (DR6), which contains 60 catalog files, 60 Segue files, and 268 raw image collection files. The sizes of each of these files varies between 5 GB and 100 GB.

We uploaded the SDSS files to several specific locations in order to better cover North America, Asia Pacific, and Europe. We then set up a web site (sdss.ncdm.uic.edu) so that users could easily obtain a Sector client application and the list of SDSS files to download. The MD5 checksum for each file is also posted on the website so that users can check the integrity of the files.

The system has been online since July 2006. During the last 2 years, we have had about 6000 system accesses and a total of 250 TB of data was transferred to end users. About 80% of the users are just interested in the catalog files, which contain files that range in size between 20 GB and 25 GB each.

Figure 4 shows the file downloading performance in an experiment of our own, where the clients are also connected to the Teraflow Testbed by 10 Gbps links. In this experiment, the bottleneck is the disk IO speed.

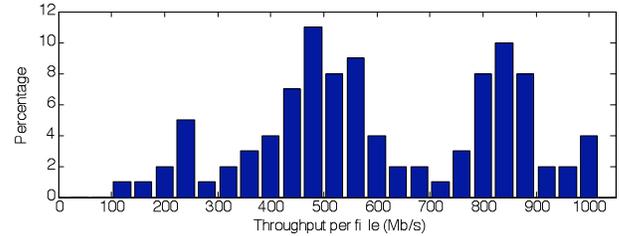

**Figure 4. File downloading performance on TFT.**

Figure 5 shows a distribution of the data transfer throughput of actual transfers to end users during the last 18 months. In most of the SDSS downloads, the bottleneck is the network connecting the Teraflow Testbed to the end user and simply using multiple parallel downloads will not help. The SDSS downloads are currently distributed as follows: 31% are from the U.S., 37.5% are from Europe, 18.8% are from Asia, and 12.5% are from Australia. The transfer throughput to users varies from 8 Mb/s (to India) to 900 Mb/s (to Pasadena, CA), all via public networks. More records can be found on sdss.ncdm.uic.edu/records.html.

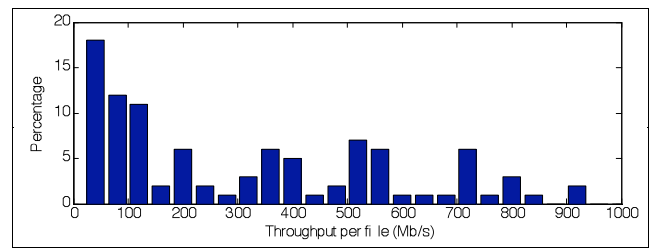

**Figure 5. Performance of SDSS distribution to end users.**

### 4.2 Terasort

We implemented the Terasort benchmark to evaluate the performance of Sphere [9, 23]. Suppose there are N nodes in the system, the benchmark generates a 10 GB file on each node and sorts the total N*10GB data. Each record contains a 10-byte key and a 90-byte value. The Sphere implementation follows the bucket sorting algorithm depicted in Figure 3.

The experimental studies summarized in Table 1 were done using the Open Cloud Testbed [20]. Currently the testbed consists of 4 racks. Each rack has 32 nodes, including 1 NFS server, 1 head node, and 30 compute/slave nodes. The head node is a Dell 1950, dual dual-core Xeon 3.0GHz, 16GB RAM. The compute nodes are Dell 1435s, single dual core AMD Opteron 2.0 GHz, 4 GB RAM, and 1 TB single disk. The 4 racks are located in JHU (Baltimore), StarLight (Chicago), UIC (Chicago), and Calit2 (San Diego).

The nodes on each rack are connected by two Cisco 3750E switches, but only a 1 Gbps connection is enabled at this time (a maximum 2 Gbps can be enabled in/out each node). The bandwidth between racks is 10 Gbps. The wide area links are provided by Cisco's C-Wave, which uses resources from the National Lambda Rail. Links from regional 10 GE research networks are used to connect the C-Wave to the racks in the testbed.

Both Sector [24] and Hadoop [19] are deployed over the 120-node (240-core) wide area system. The master server for Sector and the name node/job tracker of Hadoop are installed on one or more of the 4 head nodes. Both the Sphere client and the Hadoop client submit the job from a node in the system. This does not effect the performance since the traffic to/from the clients is negligible.

Table 1 lists the performance for the Terasort benchmark for both Sphere and Hadoop. The time is in seconds and time to generate the data is not included. Note that it is normal to see longer processing time for more nodes because the total amount of data also increases proportionally.

In this experiment, we sort 300GB, 600GB, 900GB, and 1.2 TB data over 30, 60, 90, and 120 nodes, respectively. For example in the last case, the 1.2 TB data is distributed on four racks located in four data centers across the USA. All 120 nodes participated in the sorting process and essentially all of the 1.2 TB data is moved across the testbed during the sort.

|  | Sector / Sphere | Hadoop 3 replicas | Hadoop 1 replica |
|---|---|---|---|
| UIC (1 location, 30 nodes) | 1265 | 2889 | 2252 |
| UIC + StarLight (2 locations, 60 nodes) | 1361 | 2896 | 2617 |
| UIC+StarLight+ CalIT2 (3 locations, 90 nodes) | 1430 | 4341 | 3069 |
| UIC+StarLight+ CalIT2+JHU (4 locations, 120 nodes) | 1526 | 6675 | 3702 |

**Table 1. The Terasort benchmark for Sector/Sphere and Hadoop. All time are in seconds.**

Both Sector and Hadoop replicate data for safety. The default replication strategy for both is to generate 3 replicas. Their replication strategies are different though. Hadoop replicates data during the initial writing, while Sector checks periodically, and, if there are not a sufficient number of replicas, it creates them. For this reason, Table 1 reports results for Hadoop with the replication sent to a replication factor of 1 (no replication), as well as the default replication factor of 3.

The results show that Sphere is about twice as fast as Hadoop (when Hadoop's replication factor is set to 1). Moreover, Sphere scales better as the number of racks increases (1526/1265=1.2 for Sphere vs. 3702/2252=1.6 for Hadoop).

## 5. RELATED WORK

In this Section, we describe some work related to Sector and Sphere. Sector provides some of the functionality of distributed file systems (DFS), such as GPFS, Lustre, and PVFS [13]. Distributed file systems provide the functionality of a file system on clusters of computers, sometimes, although rarely, over geographically distributed locations. While DFS may be suitable for a single organization with dedicated hardware and management, it is challenging to deploy and operate a regular DFS on loosely coupled infrastructure consisting of commodity computers, such as those used for the experimental studies described here.

On the other hand, the Google File System (GFS) [8], the Hadoop Distributed File System (HDFS) [23], and Sector are special purpose file systems. They are particularly optimized for large files, for large scanning reads, and for short random reads. In contrast to Sector, neither GFS nor HDFS were designed for nodes distributed over a wide area network.

The Sector servers that are deployed over the Teraflow Testbed and used for distributing the SDSS data provide the functionality of a content distribution network (CDN), such as Akamai [5]. Akamai is designed to distribute large numbers of relatively small files and keeps a cache at most edge nodes of its network. In contrast to Akamai, Sector is designed to distribute relatively small numbers of large files and maintains copies at several, but not all, edge nodes.

The stream-processing paradigm in Sphere is currently quite popular in the general purpose GPU programming (GPGPU) community [14]. The approach with GPGPU programming is to define a special "kernel function" that is applied to each element in the input data by the GPU's vector computing units. This can be viewed as an example of an SIMD (single instruction multiple data) style programming. Many GPU programming libraries and programming languages (e.g., Cg, sh, and Brook) have been developed. Similar ideas have also been applied to multi-core processors, including the Cell processor. For example, specialized parallel sorting algorithms have been developed for both GPU processors [9] and the Cell processor [7].

Sphere uses the same basic idea, but extends this paradigm to wide area distributed computing. Many of the GPGPU algorithms and applications can be adapted and run in a distributed fashion using Sphere. In fact, it is this analog to GPGPU that inspired our work on Sphere.

There are some important differences though: Sphere uses heterogeneous distributed computers connected by high speed, wide area networks instead of the identical ALUs integrated in a GPU. Sphere supports more flexible movement of data, but also requires load balancing and fault tolerance. Finally, the bandwidth between Sphere's SPEs, although it may be up to 10 Gbps, is not even close to the bandwidth within a GPU. Because of these differences, Sphere runs a complete processing function or program on each SPE, rather than one instruction.

One way of viewing GPGPU style programming is as a parallel programming style that gains simplicity by restricting the type of application that it is targeting. In contrast message passing systems such as MPI [25] are designed for very general classes of applications but are usually harder to program. Google's MapReduce [4] is one of the most well known examples of a system that targets a limited class of applications, but is relatively simple to program. As shown above, the Sphere system is similar to, but more general than MapReduce. Sphere is a generalization of MapReduce in the sense that it provides a simple mechanism to execute User Defined Functions (UDF) over data managed by Sector. Sphere can implement a MapReduce by using a UDF Map followed by a UDF Reduce.

See Table 2 for a summary of some of the differences between Sector/Sphere and other systems for cloud computing.

Sector/Sphere is similar to grid computing in that it aggregates distributed computing resources, but its approach is quite different. Traditional grid systems such as the Globus Toolkit [6] and Condor [15] allow users to submit multiple tasks and to run these tasks in parallel. Grid job schedulers such as Swift [16] and DAGMan [17] provide workflow services to support the scheduling of user tasks. Grid systems manage relationships among many tasks. In contrast, the Sphere client scheduler exploits data parallelism within one task. In this sense, grid computing is task oriented (multiple tasks processing one or more datasets), while Sphere is data oriented (single program processing a single dataset).

In addition, a Grid application submits a user's tasks to computing resources and moves data to these resources for computation, whereas Sector provides long term persistent storage for data and Sphere is designed to start operations as close to the data as possible. In the Sphere targeted scenarios, datasets are usually very large and moving them is considerably expensive. To summarize, Grid systems are designed to manage scarce specialized computing resources, while storage clouds, such as Sector, are designed to manage large datasets and compute clouds, such as Sphere, are designed to support computation over this data.

Finally, note that Sphere is very different from systems that process streaming data such as GATES [3] and DataCutter [1] or event stream processing systems such as STREAM, Borealis, and TelegraphCQ [26]. While Sphere is designed to support large datasets, the data being processed is still treated as finite and static and is processed in a data-parallel model. In contrast, event stream processing systems regard the input as infinite and process the data with a windowed model, sometimes with filters incorporating timing restraints in order to guarantee real time processing.

| Design Decision | GFS, BigTable | Hadoop | Sector/Sphere |
|---|---|---|---|
| Datasets divided into files or into blocks | blocks | blocks | files |
| Protocol for message passing within the system | TCP | TCP | Group Messaging Protocol (GMP) |
| Protocol for transferring data | TCP | TCP | UDP-Based Data Transport (UDT) |
| Programming model | MapReduce | MapReduce | user-defined functions applied to segments |
| Replication strategy | replicas created at the time of writing | replicas created at the time of writing | replicas created periodically by system |
| Support high volume inflows & outflows | No | No | Yes, using UDT |
| Security model | Not mentioned | None | User-level and file-level access controls |
| Language | C++ | Java | C++ |

**Table 2. A summary of some of the differences between Sector/Sphere and GFS/BigTable and Hadoop.**

## 6. CONCLUSIONS

For several years now, commodity clusters have been quite common. Over the next several years, wide area, high performance networks (10 Gbps and higher) will begin to connect these clusters. At the risk of oversimplifying, it is useful to think of high performance computing today as an era in which cycles are the scarce resource, and (relatively small) datasets are scattered to large pools of nodes when their wait in the queue is over. In contrast, we are moving to an era in which there are large distributed datasets that must be persisted on disk for long periods of time and high performance computing must be accomplished in a manner that moves the data as little as possible, due to the costs incurred when transporting large datasets.

Sector and Sphere are designed for these types of applications involving large, geographically distributed datasets in which the data can be naturally processed in parallel. Sector manages the large distributed datasets with high reliability, high performance IO, and a uniform access. Sphere makes use of the Sector distributed storage system to simplify data access, increase data IO bandwidth, and exploit wide area high performance networks. Sphere presents a very simple programming interface by hiding data movement, load balancing, and fault tolerance.

## 7. ACKNOWLEDGMENTS

The Sector/Sphere software system is funded in part by the National Science Foundation through Grants OCI-0430781, CNS-0420847, and ACI-0325013.

## 8. REFERENCES

[1] M. D. Beynon, R. Ferreira, T. Kurc, A. Sussman, and J. Saltz, DataCutter: Middleware for filtering very large scientific datasets on archival storage systems. Mass Storage Systems Conference, College Park, MD, March 2000.

[2] Fay Chang, Jeffrey Dean, Sanjay Ghemawat, Wilson C. Hsieh, Deborah A. Wallach, Mike Burrows, Tushar Chandra, Andrew Fikes, and Robert E. Gruber, Bigtable: A Distributed Storage System for Structured Data, OSDI'06, Seattle, WA, November, 2006.

[3] Liang Chen, K. Reddy, and Gagan Agrawal. GATES: A Grid-Based Middleware for Processing Distributed Data Streams. 13th IEEE International Symposium on High Performance Distributed Computing (HPDC) 2004, Honolulu, Hawaii.

[4] Jeffrey Dean and Sanjay Ghemawat, MapReduce: Simplified Data Processing on Large Clusters, OSDI'04: Sixth Symposium on Operating System Design and Implementation, San Francisco, CA, December, 2004.

[5] J. Dilley, B. Maggs, J. Parikh, H. Prokop, R. Sitaraman and B. Weihl, Globally Distributed Content Delivery, IEEE Internet Computing, page 50-58, September/October 2002.

[6] I. Foster. Globus Toolkit Version 4: Software for Service-Oriented Systems. IFIP International Conference on Network and Parallel Computing, Springer-Verlag, LNCS 3779, pages 2-13, 2005.

[7] Bugra Gedik, Rajesh Bordawekar, Philip S. Yu: CellSort: High Performance Sorting on the Cell Processor. VLDB 2007: 1286-1207.

[8] Sanjay Ghemawat, Howard Gobioff, and Shun-Tak Leung, The Google File System, 19th ACM Symposium on Operating Systems Principles, 2003.

[9] Naga K. Govindaraju, Jim Gray, Ritesh Kumar, Dinesh Manocha, GPUTeraSort: High Performance Graphics Coprocessor Sorting for Large Database Management, ACM SIGMOD 2006.

[10] Jim Gray. Sort benchmark home page. http://research.microsoft.com/barc/SortBenchmark/.

[11] Yunhong Gu, Robert L. Grossman, *UDT: UDP-based Data Transfer for High-Speed Wide Area Networks*. Computer Networks (Elsevier). Volume 51, Issue 7. May 2007.

[12] Yunhong Gu, Robert L. Grossman, Alex Szalay and Ani Thakar, Distributing the Sloan Digital Sky Survey Using UDT and Sector, Proc. of e-Science 2006.

[13] W. T. C. Kramer, A. Shoshani, D. A. Agarwal, B. R. Draney, G. Jin, G. F. Butler, and J. A. Hules, Deep Scientific Computing Requires Deep Data, IBM Journal of Research and Development, March 2004.

[14] John D. Owens, David Luebke, Naga Govindara Ju, Mark Harris, Jens Kruger, Aaron E. Lefohn, and Timothy J.Purcell, A Survey of General-Purpose Computation on Graphics Hardware, Eurographics 2005, pages 21-51, 2005.

[15] Douglas Thain, Todd Tannenbaum, and Miron Livny, "Distributed Computing in Practice: The Condor Experience" *Concurrency and Computation: Practice and Experience*, Vol. 17, No. 2-4, pages 323-356, February-April, 2005.

[16] Y. Zhao, M. Hategan, B. Clifford, I. Foster, G. von Laszewski, I. Raicu, T. Stef-Praun, and M. Wilde, Swift: Fast, Reliable, Loosely Coupled Parallel Computation IEEE International Workshop on Scientific Workflows 2007.

[17] The Condor DAGMan (Directed Acyclic Graph Manager), http://www.cs.wisc.edu/condor/dagman.

[18] Amazon Web Services, http://www.amazon.com/aws.

[19] Hadoop, http://hadoop.apache.org/core.

[20] The Open Cloud Testbed, http://www.opencloudconsortium.org.

[21] Teraflow testbed, http://www.teraflowtestbed.net.

[22] Robert L Grossman and Yunhong Gu, Data Mining Using High Performance Clouds: Experimental Studies Using Sector and Sphere, Proceedings of The 14th ACM SIGKDD International Conference on Knowledge Discovery and Data Mining (KDD 2008), ACM, 2008

[23] Dhruba Borthaku, The Hadoop Distributed File System: Architecture and Design, retrieved from lucene.apache.org/hadoop, 2007.

[24] Sector is available from sector.sourceforge.net.

[25] William Gropp, Ewing Lusk and Anthony Skjellum, Using MPI: Portable Parallel Programming with the Message Passing Interface, Second edition, MIT Press, 1999.

[26] Brian Babcock, Shivnath Babu, Mayur Datar, Rajeev Motwani, and Jennifer Widom, Models and issues in Data Stream Systems, In Proceedings of the Twenty-First ACM SIGMOD-SIGACT-SIGART Symposium on Principles of Database Systems, PODS 2002, ACM, New York, pages 1-16.